# Usability Meets Instant Gratification on the Semantic Web – Structure Helps a Wiki Navigate

David Aumueller

Department of Computer Science
University of Leipzig, Augustusplatz 10/11, 04103 Leipzig, Germany
david@informatik.uni-leipzig.de

**Abstract.** This paper presents a semantic wiki prototype application named SHAWN that allows structuring concepts within a wiki environment. To entice the use of Semantic Web technologies applications need to offer both high usability and instant gratification. Concept creation is exceptionally easy in SHAWN since metadata as well as plain text is entered within a single edit box on each wiki page in a self-explaining fashion. The entered data is immediately used for rendering sophisticated navigational means on the wiki. By editing simple wiki pages ontologies emerge.

## 1 Introduction

The Semantic Web, a vision of the future Web, should give information on the current Web "well-defined meaning [to better enable] computers and people to work in cooperation" [1]. With not even syntactically valid web sites one cannot expect their authors to write semantically marked up pages. It is the ease of publishing which made the Web the most successful type of media in the last decade. Its popularity emerged due to the relaxed interpretation of incomplete and invalid markup by web browser applications. Authors were instantly gratified by seeing their content published online.

The approach presented here of a semantic wiki application wants to pursue those steps by allowing to easily add structure and semantics to web pages within a wiki environment and by instantly exploiting that information both for navigational aids within the wiki web as well as creating new information via simple reasoning.

### 1.1 Motivation

The common approach to creating semantically rich data, i.e. information annotated with metadata, is to export RDF (Resource Description Framework) from existing well-structured data. RDF consists of subject-predicate-object triples that state specific facts about resources or concepts, e.g. "[Shakespeare] <isAuthorOf> [Hamlet]", whereby subject, predicate, and object (if not a literal) are identified via URIs.

This top-down approach of creating semantic markup prevails due to the fact that the opposite approach of creating semantically rich data bottom-up is tedious. To bridge the gap between the unstructured and structured world various ideas exist.

The Mangrove project [2], [3] aims to overcome the obstacles of semantic markup by defining a small set of elements that are to be used to tag data such as people's names, contact information, and publications. A tool is offered to support annotation. The Mangrove search engine "understands" this extended markup to support semantic queries. The need for such bottom-up approaches that "cross the structure chasm" is also expressed in [4]. Authoring unstructured information is straightforward being the natural way of content creation.

The approach of the prototype presented in this paper – called SHAWN – resembles both bottom-up and top-down at the same time, meeting half-way, by bridging the gap between unstructured and structured data as it allows easy content creation from the bottom but at the same time exporting rich metadata from the top and thus follows also Tim Berners-Lee's demand[1] in his recent WWW2004 talk [6]. The wiki concept is accepted by one of the largest communities on the Web – Wikipedia.org would not be that successful and could not have published almost half a million articles in the English edition within the four years of its existence. Thus, I think the wiki way [7] is the way to entice ordinary people onto the Semantic Web via instant gratification [2].

### 1.2 Related Systems

Wiki engines are a popular field of experimentation which results in an abundance of different implementations. Well known wiki webs and broadly used engines include: WikiWikiWeb[2], UseMod.com, MoinMoin[3], Wikipedia.org's MediaWiki, TWiki.org, JSPWiki.org, and ZWiki.org.

These engines implement the *backlinks* mechanism, which allows the user to request a list of wiki pages that link to the current page. In particular, this is used for categorization of wiki pages in that a special wiki page is created to resemble a specific category and to each page belonging to this category a link to this category page is created. The list of backlinks from the category page then shows all the wiki pages that link to the category and thus belong into this category. MediaWiki engine implements a categorization mechanism since just recently. The ZWiki engine e.g. supports hierarchical classification by merely specifying one parent page within a HTML form field for that purpose.

---

[1] Tim Berners-Lee "pushed developers to start using RDF and triples more aggressively. In particular, he wants to see existing databases exported as RDF, with ontologies created adhoc to match the structure of that data. Rather than using PHP scripts only to produce HTML, he suggested, create RDF as well. Then, when all of the RDF is aggregated, apply rules and see what happens." [5] Ford, P.: Berners-Lee Keeps WWW2004 Focused on Semantic Web. Online: http://www.xml.com/pub/a/2004/05/20/www-timbl.html

[2] Portland Pattern Repository Site and Software <http://c2.com/cgi/wiki?WikiWikiWeb>

[3] MoinMoin is a Python Clone of WikiWiki. <http://sourceforge.net/projects/moin>

Worth mentioning also is the OpenGuides.org wiki engine which fulfils the purpose of a cityguide listing possible spots of interests annotated with their geographical position to support querying the wiki for other places that are locally near that spot. This metadata is stored as RDF annotations that have to be entered into very specific HTML form fields on each wiki page.

Generally, two main approaches are conceivable for representing RDF triples within a wiki environment. Firstly, one concept is denoted by an entire wiki page with its page name resembling the RDF subject. Secondly, marked sections of wiki pages each denote a different concept, i.e. one page could contain multiple subjects. Both PlatypusWiki [8] and the herein presented prototype adhere to the first option – one page, one concept.

The PlatypusWiki is a promising effort to implement a general RDF wiki engine that supports the use of RDF and OWL (Web Ontology Language) vocabularies to represent metadata and relations between wiki pages. Content and metadata have to be edited separately though. Although entering RDF triples is done via specific HTML form fields for subject, predicate, and object, the engine does not (yet) enforce any constraints – inconsistencies are to be sorted out by the community.

### 1.3 A glimpse of SHAWN

The here presented prototype SHAWN is a lightweight wiki engine that allows creating and editing web pages (wiki pages) within the site without the need of specialists' knowledge but with far reaching possibilities. It inhibits features that add both levels of structure and usability to the wiki paradigm which are to my knowledge not to be found in any related projects. These features include the easy editing of both content and metadata at the same time as well as immediately gratifying the user with enhanced navigational means, such as breadcrumbs and forwardlinks – all based on the entered data. Structure is established by entering simple field-value pairs (properties) that are interpreted as typed hyperlinks or relationship types. The emergent graph of wiki pages resembles a rich model that can be exported as Semantic Web ontology.

## 2 Use Cases

The most prominent wikis – the original WikiWikiWeb by Ward Cunningham, and the Wikipedia – are communities that maintain pages about arbitrary concepts and about their users or contributors themselves. Suppose the users state on their pages such facts as what skills or interests they have, where they live, and whom they know personally on the wiki. With simple inference a user could determine e.g. those persons that she might ask for help in specific matters or just might want to meet to share common interests. Taking also a (transitive) "knows" property into account the result could be granulated down to those people the questioner knows personally or could get to know via other people.

This corresponds with similar purposes online social network sites such as Friendster.com, Orkut.com, and OpenBC.com are designed for. Users log on to these web sites to communicate with friends and get to know further people via already established links for either specific business interests or leisure activities. The supporters of the Friend of a Friend (FOAF) vocabulary try to accomplish similar goals using a decentralized approach by encouraging people to publish RDF documents on their homepage that describe their interests and list some friends. With this variety of online social networks users may have distributed their social network across various online social network sites and know some people on Friendster, some on Orkut, others on OpenBC, and again others that have some FOAF data published. Thus, one might wish to integrate all the information relevant for oneself accessible under one roof – a perfect scenario for the SHAWN wiki application in private use, easily accessible from anywhere on the Web. Using the wiki paradigm as PIM (Personal Information Management) got also suggested by Leuf and Cunningham in [7]. With its tiny size the wiki application will never compete with a Goliath sized PIM application such as Haystack [9], though.

This kind of personal semantic notebook with address book and social network could also be enriched from time to time to hold facts such as how one got to know each other, e.g. via whom or via which location or event – using this kind of "got to know by" relation in a transitive manner creates an interesting graph of people and space relationships. Other properties to collect include e.g. dates when one has last met each other, has last spoken to, or simply the date of birth. Such dates subsequently could serve as triggers for reminders.

The available contact data could be exported to various address book formats, such as LDIF or vCard or when set up in conjunction with a Lightweight Directory Access Protocol (LDAP) server accessed directly from within one's favourite email client. Further, storing received and sent emails on the personal wiki creates a navigable email categorization connecting incoming mails with sent mails. The emergent graphs of conversations could be further annotated with notes and related attachments, e.g. publications by this contact person. Categorizing and cross linking such bibliographic data would give a perfect community project by itself for making bibliographical entries of scientific publications accessible similar to the freedb.org project where the community enters song track titles from audio CDs accessible for others via their favourite CD player application – for publications BibTeX import/export and Z39.50[4] protocol support would be the appropriate methods of access. The Bibster [10] project e.g. similarly tries to accomplish the interchange of such publication metadata within a peer-to-peer (P2P) environment.

Imagine a combination of all the above mentioned scenarios – all the people you know, their contact data, the online conversations you had with them, as well as their publications. Now, suppose you would like to plan a trip or sabbatical where you wish to meet as many researchers as possible who share your interests. Combining properties or facts such as LivesIn, AuthorOf, CoversTopic, and InterestsIn could finally infer a list of people matching your research interests and at the same time all live relatively close together.

---

[4] A protocol used e.g. in reference toolkits for word processors to search and retrieve bibliographic entries from libraries.

As last use case SHAWN may serve as online learning experience to help teachers and pupils alike in structuring and/or understanding complex pieces of information. In a literature class e.g. the teacher might ask the students to model the relationships between the characters in a Shakespearean play.

## 3 The SHAWN prototype

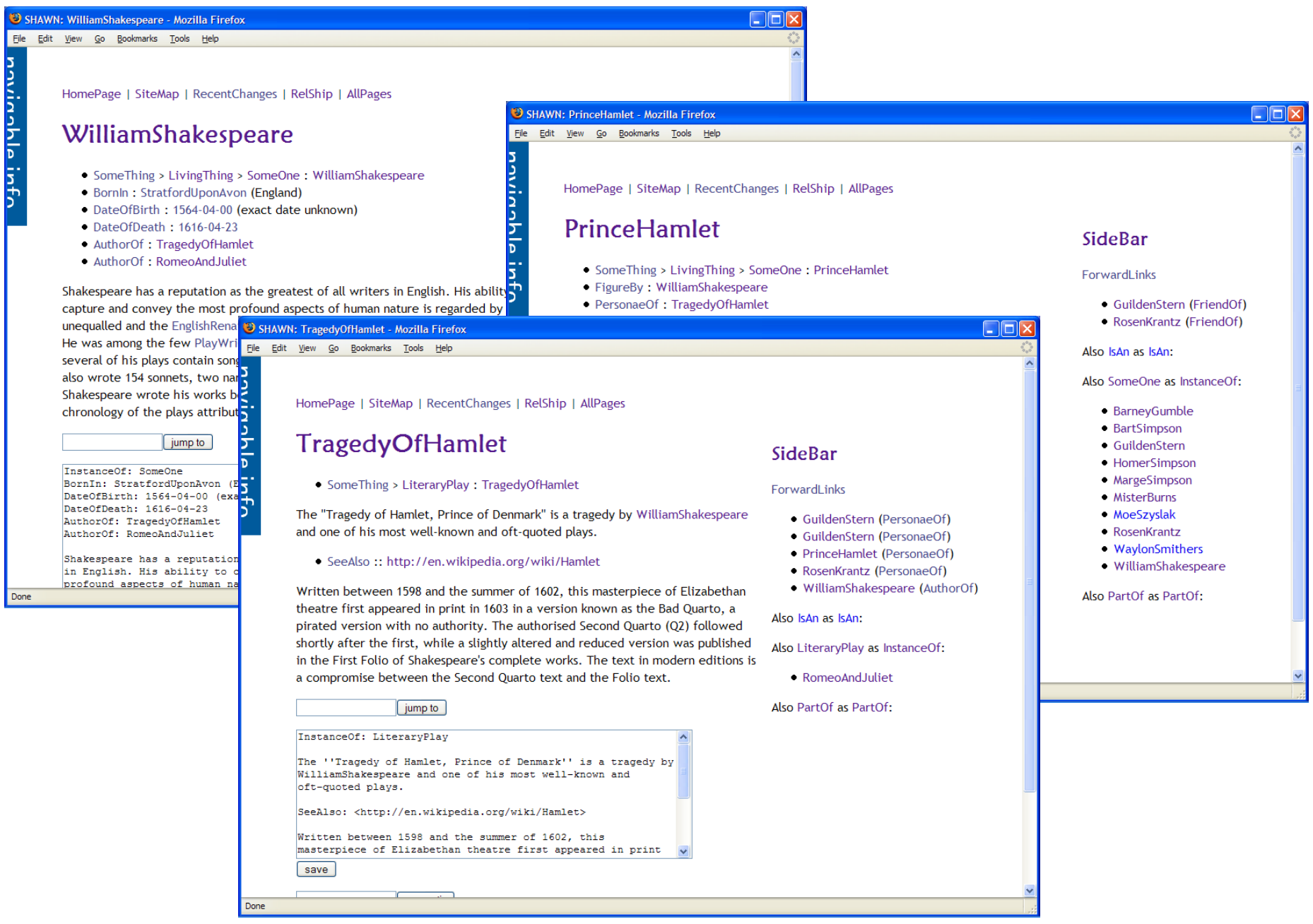

**Fig. 1.** Screenshots of related wiki pages within the SHAWN prototype

### 3.1 Usability in both Content Creation and Navigation

The ease of putting content online on the Web culminates in wikis, where the content of each page can be edited on the page itself. Adapting this approach the SHAWN prototype allows both entering structure information or metadata and further free text describing the concept – all within the same edit box on the wiki page which resembles the concept. With typical field value pairs one creates RDF-like triples (subject-predicate-object) with subject being the page (name) itself, property the field, and object/literal the assigned value. This approach also inhibits no redundancy as it is the case in annotating existing HTML pages since parts of the information need to be duplicated as machine-readable code [11]. Here, field-value pairs are only entered once and according RDF triples are produced from them automatically. Annotating existing HTML pages would mean to generate redundancy.

The structural information is instantly used to enhance navigation. Navigational aids as demanded by Jacob Nielson [12] for maximum usability of websites include answers to the questions “Where am I” and “Where can I go”. The first gets credited within SHAWN with so called *breadcrumbs* that show the path to the current page from the root of the site or concept (following specific transitive relationship types). The second is resembled by what I will call *forwardlinks.* These display the page names of those pages that link to the current page via arbitrary relationship types entered within their concept wiki page. Thus, instant gratification shows the user the value of structuring the data.

### 3.2 Instant Gratification

Every concept, be it a subject or predicate in the sense of RDF, is resembled as wiki page. The *forwardlinks* show all to the current concept related pages. If a concept is used as predicate, i.e. when a wiki page takes the role of a relationship type, the wiki engine will list all triples that use this predicate. With this mechanism the user immediately sees all the other concepts that share the same property, showing also the objects or literal values used therein. For example, a model containing some persons as concepts involving a relationship type “LivesIn” would, on this page, show all other persons that have this property together with the actual values of these properties, i.e. the whole triples.

Since it is tremendously easy to add arbitrary facts to one concept (merely entering field value pairs in the edit box of the wiki page), the user can add anything that comes into her mind regarding a concept and can make up any kind of field value pairs (i.e. relationship types or predicates). In a later stage, these could further be typed using an is-a relationship type to build hierarchies of relationship types. This will reduce the structural entropy/mess of early creative sessions and a rich ontology will emerge by itself.

Instantly gratifying the user while authoring the content conflicts with enforcing possible integrity constraints, though. This has to be deferred to a later stage, e.g. when the resulting ontology exported OWL ought to be used in full fledged editors, such as Protégé [13], Swoop [14], or pOWL [15]. McBride [16] points out that it has to be accepted that flawed information models will be prevalent on the Semantic Web.

### 3.3 Flexibility of the Implementation

The prototype currently is implemented in about 500 lines of Perl, running on an Apache web server. Each wiki page is stored on the server as plain text file. After processing potential special wiki commands, the resulting text is transformed to XHTML using MarkDown [17]. Common “CamelCase” words act as wiki links and thus as general concepts; so called freetext wiki links (e.g. masking wiki links by enclosing with double square parenthesis) could be easily implemented as well. To enter the semantic metadata (property-value pairs) no special markup is needed. Field value pairs are written on the wiki edit box simply as in “property: value”. These pairs get parsed from each page and interpreted as subject-predicate-object triples for further use in support of navigation at various spots on a wiki page.

Navigational aids are located at the top (fixed links such as HomePage, AllPages, as defined in a special wiki page called GotoBar) and on the right-hand side (sidebar) where links dependent on the current page are displayed, especially the aforementioned forwardlinks. The sidebar is indeed a wiki page by its own (SideBar) and is to contain special wiki commands that get transformed to lists or trees of links to semantically related wiki pages. Those operators can be used anywhere on any wiki page. Placing them in the special sidebar page makes them being executed and the results displayed for each page accordingly. By default, all pages which contain the current page as object of any property get listed (including the property) as forwardlinks and e.g. all pages which are of the same type as the current page get listed in the sidebar, as well. Thus, breadcrumbs make it easy to go up in the hierarchy; forwardlinks open way to venture deeper into the site or into more specialized concepts without getting lost.

Another wiki command outputs all the known triples when used on wiki pages that resemble a relationship type such as InstanceOf or LivesIn. Classifying those relationship type wiki pages enhances navigation the same way as described above: A list of available relationship types is shown in the sidebar since those are pages that share the same type.

Only few properties are intrinsically known to the wiki engine for specific rendering of the wiki pages. These are at the moment “TypeOf” and “InstanceOf”. The transitive type-of property and the instance-of property are used to render breadcrumbs so that the user knows exactly where her page or concept is located in context. As usual, the breadcrumbs path gets concatenated via greater than symbols (‘>’); instances get appended to the breadcrumbs via the colon notation (‘: instance’), common for instances. A special “PartOf” type could be used to build larger wiki pages consisting of multiple pages which then could be e.g. exported to XML DocBook. The behaviour (e.g. transitivity) of these special relationship types will be defined on their wiki page in the future.

### 3.4 Visualization and further Navigational Aids

As visualization of either trees of specific relationship types or whole graphs of the complete semantic wiki structure the prefuse toolkit [18] was chosen. This toolkit contains various layout routines to display hierarchical and graph data in an interactive fashion that allows changing focus of nodes in context. Integrated as Java applet into the wiki environment it further enhances navigation and offers a more complete overview of complex structures. Planned to implement are controls to change visibility and appearance (width and colour) of each relationship type edge to filter the visualization of large graphs.

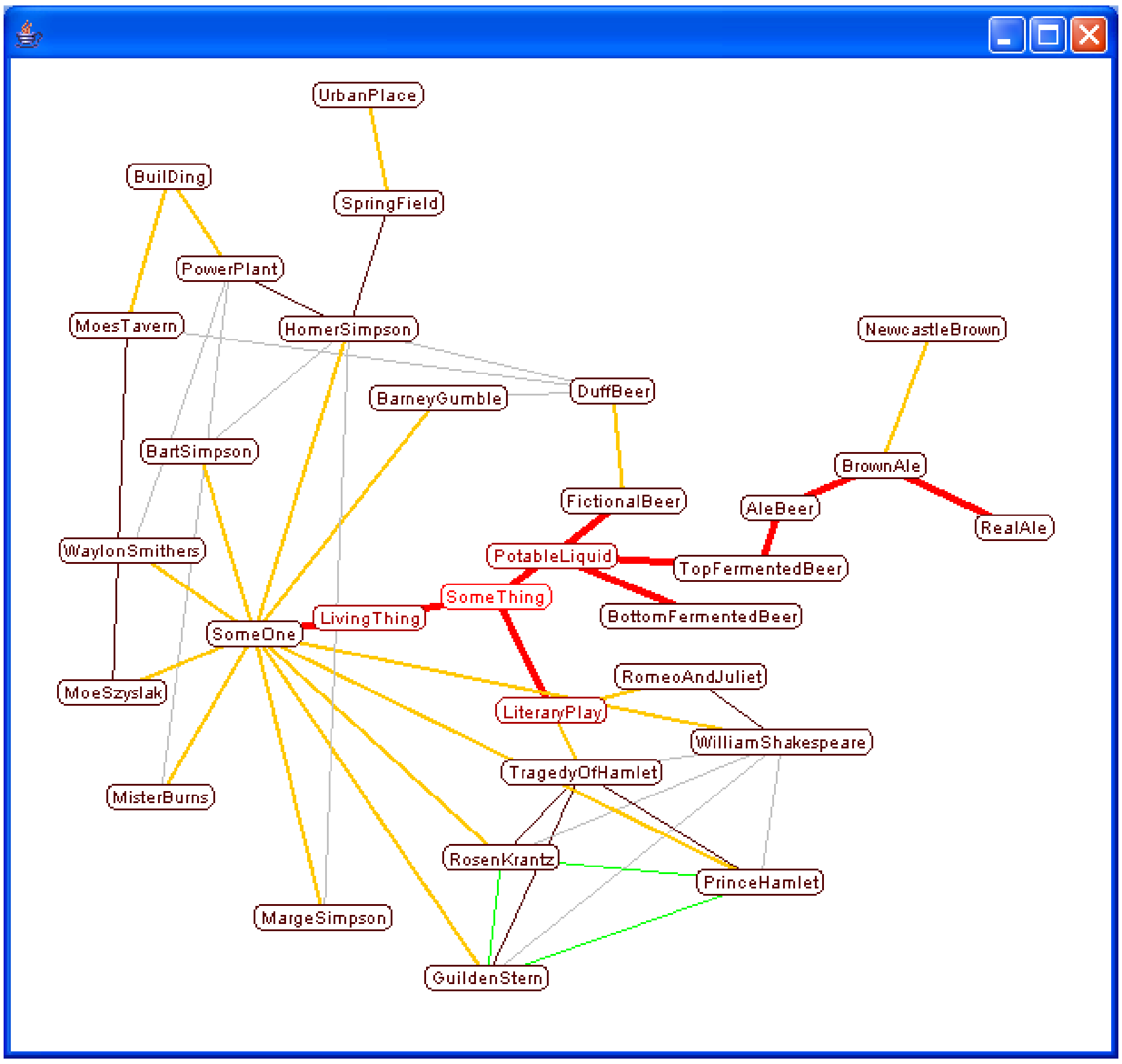

**Fig. 1.** Interactive graph of wiki pages as rendered by the prefuse toolkit (radial layout)

## 4 Discussion and Future Integration

The wiki approach is a very flexible way of structuring pieces of information. For a start, this entices feeding lots of data into the wiki. Later on, the pieces can easily be rearranged by merely changing some field value-pairs or doing site wide search-and-replaces. An important aspect in ontology creation within the Semantic Web is to adhere to given structures and use vocabularies already available. When referring to a concept that is already defined by some ontology or vocabulary it should be used or refined instead of re-invented to keep heterogeneity and integration efforts low. The SHAWN prototype supports this by merely stating equality of concepts with concepts defined in external resources, i.e. URIs. For the RDF/OWL export, a special relationship type “SameAs” may be used to denote that e.g. a concept “SomeOne” denotes the same concept as <http://xmlns.com/foaf/0.1/person>, and the export module will thus replace references to SomeOne with the URI from the FOAF vocabulary. The RDF generation will be run each time a page got edited to keep the "RDF behind" up-to-date and searchable for Semantic Web crawlers.

Nevertheless, the SHAWN wiki is not meant to be a full fledged RDF/OWL editor. The flexible approach of this prototype does not enforce any semantics.

### 4.1 Semantic Retrieval through Inferences

By exporting the emergent triples to RDF in its XML or N3 notation, using inference engines in conjunction with some rules further triples can be deduced. As inference processor Tim Berner-Lee's CWM [19] or Sean B. Palmer's EEP[5] are easily deployed. More complex solutions, e.g. deductive databases such as XSB[6] or special language implementations such as TRIPLE [20] will be scrutinized for its adoptability in the wiki context.

### 4.2 Scalability Issues

At the moment, each wiki page is stored in one plain text file. To be scaleable an underlying database will be needed. Taking a standard approach with relational databases into account, the following concerns have to be considered thinking about an appropriate relational schema. Wikis being maximally flexible data repositories, the database schema for storing the information has to be very simplistic. The naïve approach is to store all the triples in one relation; even the whole page content can be seen as literal value within an RDF triple "[WikiPage] <hasContent> content". The relationship data (field-value pairs) would need to be extracted upon saving (creating/updating) a wiki page and then put into the triple store. Still, this relation approach does not inhibit any typing, i.e. all attributes contain character data. Numerical comparisons (e.g. a query incorporating specific points in time[7]) would need to try casting all strings to numeric values. Hence, dynamic creation of database relations that are typed accordingly should be considered. For instance, triples covering "[Person] <DateOfBirth> [1948-03-20]" the relation should be typed varchar, varchar, date. Further suggestions of how to store RDF in relational databases [21] were collected by Sergey Melnik.

## 5 Conclusion

The prototype of a semantic wiki application presented here under the name SHAWN shows how a very simplistic approach to structuring data could succeed. It is the ease of publishing content on the Web that needs to be pursued further in the Semantic Web. Already little effort needs to be credited immediately as it is the case in editing general wiki pages. Whether it is only correcting typos or adding valuable content to a wiki, the result is instantly visible to the contributor and the whole community. To entice lots of users onto new technologies instant gratification is of vital importance. With the straightforwardness of entering structural metadata to SHAWN wiki pages the user gets instantly gratified by additional navigational links resembling the structure of a growing model. This may even elicit further ideas to be entered in the wiki. The underlying semantics of the emergent ontology offer the user all the possibilities for her data existing and yet-to-come Semantic Web technology has to offer.

---

[5] Eep3: CWM Clone and SW API <http://infomesh.net/2002/eep3>

[6] Logic Programming and Deductive Database System <http://xsb.sourceforge.net>

[7] E.g.: "Who was born the same year as Shakespeare?", "Whom of my close friends did I not yet meet this year?"